\documentclass[]{aa}
\usepackage{natbib}
\usepackage{bm}
\usepackage{svg}
\usepackage{ulem}
\usepackage{graphicx}
\usepackage{txfonts}
\newcommand{\cm}{\mbox{\,cm}}

\newcommand{\degK}{\,{}^\circ\mbox{K}}

\newcommand{\erg}{\mbox{\,erg}}

\newcommand{\kms}{\mbox{\,km s}^{-1}}

\newcommand{\muG}{\,\mu\mbox{G}}
\newcommand{\Myr}{\mbox{\,Myr}}
\newcommand{\pc}{\mbox{\,pc}}
\newcommand{\pcc}{\mbox{\,cm}^{-3}}

\usepackage{xcolor}
\begin{document}

   \title{The role of shocks and the velocity gradient in the relative orientation of the magnetic field and dense gas clouds}
   
   \titlerunning{Magnetic field relative orientation in dense gas clouds}
   \authorrunning{Granda-Mu\~noz et al.}
  
   %\subtitle{I. Overviewing the $\kappa$-mechanism}
   \author{Guido Granda-Mu\~noz,\inst{1,2} Enrique V\'azquez-Semadeni,\inst{2} \and Gilberto C. G\'omez \inst{2}}
\institute{ Departamento de Ciencias, Facultad de Artes Liberales, Universidad Adolfo Ib\'a\~nez, Av. Padre Hurtado 750, Vi\~na del Mar, Chile \\
\email{guido.granda@edu.uai.cl}
\and
Instituto de Radioastronom\'ia y Astrof\'isica, Universidad Nacional Aut\'onoma de M\'exico, Apdo. Postal 3-72, Morelia, Michoac\'an 58089, M\'exico \\
\email{e.vazquez@irya.unam.mx,g.gomez@irya.unam.mx}
}
\date{Received 2024 May 22; accepted 2025 January 14}

% \abstract{}{}{}{}{}
% 5 {} token are mandatory

  \abstract
  {
 Magnetic fields are known to exhibit different relative orientations with  density structures in different density regimes. However, the physical mechanisms behind these relative orientations remain unclear.
  }
  {We investigate the role of the flow features on the relative orientation between the magnetic field and cold neutral medium (CNM) clouds, as well as that of molecular clouds (MCs)  as a
corollary.}
  {We performed three- and two-dimensional (3D+2D) magnetohydrodynamic (MHD) simulations of warm gas streams in the thermally bistable atomic interstellar medium (ISM) colliding with velocities of the order of the velocity dispersion in the ISM to form CNM clouds. In these simulations, we followed the evolution of magnetic field lines to identify and elucidate the physical processes behind their evolution.}
  {The collision produces a fast MHD shock, as well as a condensation front roughly one cooling length behind it, on each side of the collision front. 
  A compressive, decelerating velocity field arises between the shock and the condensation fronts, and a cold dense layer forms behind the condensation front. The magnetic field lines, initially oriented parallel to the flow direction, are perturbed by the fast MHD shock, across which the magnetic field fluctuations parallel to the shock front are amplified. The downstream perturbations of the magnetic field lines are further amplified by the compressive downstream velocity gradient between the shock and the condensation front caused by the settlement of the gas onto the dense layer. This process causes the magnetic field to become progressively aligned with the dense layer, leading to the formation of a shear flow around it, due to the field's backreaction on the flow. By extension, we suggest that a tidal stretching velocity gradient, such as that produced in gas infalling into a self-gravitating structure, must straighten the field lines along the accretion flow, orienting them perpendicular to the density structures. We also find that the initially super-Alfv\'enic upstream flow becomes trans-Alfv\'enic between the shock and the condensation front, and then sub-Alfv\'enic inside the condensation. Finally, in 2D simulations with a curved collision front, the presence of the magnetic field inhibits the generation of turbulence by the shear around the dense layer.}
  {Our results provide a feasible physical mechanism for orienting the magnetic field parallel to CNM clouds through the action of fast MHD shocks and compressive velocity fields.}


   \maketitle
%
%________________________________________________________________
\begin{nolinenumbers}
\section{Introduction}
Studying the role of magnetic fields in the formation and evolution of atomic and molecular clouds has been an important research topic for both observational and theoretical astronomy. Magnetic fields are thought to be an important ingredient in the dynamics of the interstellar medium (ISM), providing a possible support mechanism against gravitational collapse, and guiding the gas flow in the surroundings of filamentary structures, among many other effects \cite[e.g.,][]{role_of_b_2011,girichidis_2020,pattle_2023}. In addition, a tight correlation between the relative orientation of the magnetic field and cold neutral medium (CNM) clouds has been identified in the last decades. For example, \citet{mcclure_griffiths_2006} reported that CNM clouds detected by HI self-absorption in the Riegel-Crutcher cloud are aligned with 
the local magnetic field orientation.
More recently, \citet{clark_2014} found that HI fibers, which are  long, thin, dense structures identified in HI emission using the rolling Hough transform (RHT), are aligned with the plane-of-the-sky magnetic field measured using thermal dust emission. \citet{clark_2015} also observed a similar alignment with atomic hydrogen structures detected by HI emission and molecular structures traced by dust.

In higher density environments, observations of the plane-of-the-sky magnetic field, detected using polarized thermal dust emission, have found that as the column density of the medium increases, a transition from parallel to perpendicular relative orientation with the elongated density structures occurs. This is seen  in the transition from the atomic to the molecular gas \citep{plank_2016_dust_align} as well as in the interiors of molecular clouds \citep[MCs,][]{plank_2016_mc_align}. \citet{plank_2016_mc_align} found that the relative orientation of the projected magnetic field and dust filaments changes from parallel to perpendicular when sampling higher density regions in nearby MCs.
Moreover, \citet{skalidis_2022} studied the role of the magnetic field in the transition between $\mathrm{H\MakeUppercase{\romannumeral 1} -H_{2}}$, using multiple tracers to investigate the gas properties of Ursa Minor. They found that turbulence is trans-Alfv\'enic and that the gas probably accumulates along magnetic field lines, generating overdensities where molecular gas can form.

However, the origin of these relative orientations remains unclear, as most of the observational evidence refers to spatial and orientation correlations without a clear understanding of the causality involved. Therefore, it is crucial to understand the interplay between the precursors of MCs and magnetic fields. Since CNM clouds are thought to constitute the primordial place for the early stages of the evolution of MCs -- and eventually star-forming regions \citep[e.g.,][]{Hennebelle_1999,Hennebelle_2000,koyama_2002, audit_thermal_2005, VZ_2006,heitsch_birth_2006,heiner_2015,seifried_2017} -- understanding the alignment mechanism of magnetic field lines in CNM clouds is relevant in elucidating the role of magnetic fields in the formation of MCs and star formation.

To study this correlation statistically, \citet{Soler_2013} proposed the histogram of relative orientations (HRO) and
found that the relative orientation of the magnetic field and isodensity contours changes from parallel to perpendicular in a high-magnetization ($\beta=0.1$) computer-simulated MCs. More recently, \cite{kortgen_2020} studied the role of the nature of turbulent forcing in the change of relative orientation, finding that it only mildly affects and that this transition occurs only for gas with a dynamically significant magnetic field. 

The relative orientation of the magnetic field with CNM clouds is arguably understood to be intrinsically related to the formation mechanism of CNM clouds. The formation of CNM clouds by colliding flows with pure hydrodynamic was studied in \citet{heitsch_birth_2006}. These authors performed simulations of colliding warm neutral gas streams, including the multi-phase nature of the ISM  and without considering gravity or magnetic fields. They discussed three important instabilities that might play a role in the formation of CNM clouds: the thermal instability (TI), the Kelvin-Helmholtz instability (KHI), and the non-linear thin-shell instability (NTSI). They concluded that these instabilities break up the coherent flows, seeding small-scale density perturbations necessary for gravitational collapse and subsequent star formation.

The formation of CNM clouds and the origin of the relative orientation of the magnetic field with density structures have been extensively studied both numerically and analytically. The first point to notice is that it is well known that the formation of CNM clouds out of the warm neutral medium (WNM) in the galactic ISM requires strong cooling in addition to the presence of shocks. Such cooling often causes TI \citep{field_1965, Pikelner68, Field+69} and this generally produces substantial density jumps, without the need for very highly supersonic flows \citep{VS+96,VZ_2006}. The CNM clouds formed in this way are thought to be the progenitors of MCs, as they continue to accrete gas and eventually become sufficiently self-shielded to form molecules and become dominated by self-gravity \citep{Hartmann_2001, Glover_ML07, VS_2007, Heitsch_2008}.

\citet{Hennebelle_2013} studied the origin of the elongation of non-self-gravitating CNM clouds in a turbulent, thermally unstable warm medium, concluding that it is caused by the stretching induced by turbulence because the clouds are aligned with the strain. Moreover, the author also found that the Lorentz force tends to confine the filamentary CNM clouds.  \citet{Inoue_2016}, in agreement with the previous work, found that the strain is also the origin of the magnetic field alignment with fibers in HI clouds formed in a shock-compressed layer using simulations resembling the local bubble. Recently, \cite{seifried_2020}  investigated the relative orientation of magnetic field and three-dimensional (3D) and two-dimensional (2D) projected structures using synthetic dust polarization maps. They found that, in agreement with the observations, the magnetic field changes its orientation from parallel to perpendicular at $n \approx 10^{2}-10^{3} \cm^{-3}$ in regions where the mass-to-flux ratio has values close to or below 1. In addition, they found that projection effects, due to the relative orientation between the cloud and the observer, affect the measurements. More recently, \citet{Gazol_2021} studied the morphology of CNM clouds in forced magnetized and hydrodynamical simulations, finding that the presence of a magnetic field increases the probability of filamentary CNM clouds.

The change in relative orientation between the magnetic field and the elongation of the density structures was studied by \citet{Soler_2017}, based on an analytic approach that involves introducing an equation to express the evolution of the quantity $\cos \phi = {\bm B} \cdot \nabla \rho/|{\bm B}| |\nabla \rho|$, where ${\bm B}$ is the magnetic field vector and $\nabla \rho$ is the density gradient. They found that the parallel or anti-parallel (respectively, $\phi=0^{\circ},180^{\circ}$) and perpendicular ($\phi=90^{\circ}$) configurations are equilibrium points of this equation. Therefore, they argued that the system tends to evolve to these $\phi$ values.

However, the above-mentioned studies have all been based on indirect analyses and measurements. For example, \citet{Inoue_2016} investigated the dependence of the alignment on the relative orientation between the magnetic field and the shock propagation direction, while \citet{seifried_2020} investigated the dependence on the background density. \citet{Hennebelle_2013} investigated the formation mechanism of filaments, finding that they are due to the strain rather than to shocks, and \citet{Soler_2017} investigated the dependence of the relative orientation between the magnetic field and the density structures on the sign of the coefficients in the derived evolution equation for $\cos(\phi)$. To our knowledge, the simultaneous temporal evolution of the density, velocity, and magnetic fields, along with the notion of how this evolution leads to the specific alignment over time and how it is related to the flow features such as  shocks, condensation fronts, and velocity gradients, has not been documented to date.

An important consequence of the supersonic nature of the collisions between gas streams that can form CNM clouds is the formation of shocks. Indeed, to trigger TI in the thermally stable WNM, a nonlinear (supersonic) compression is necessary, throwing the gas out of thermal equilibrium and causing a far-from-equilibrium evolution towards the CNM \citep{Koyama_2000, Kritsuk_2002}. Thus, the supersonic compression generates a shock, and a condensation front, one cooling length behind it. The gas flows across the latter to ultimately settle into the dense CNM layer, which is back in thermal equilibrium \citep{VZ_2006}. 

In this paper, we investigate the detailed simultaneous evolution of the magnetic field, as traced by the evolution of the field lines and the field's amplitude, as well as of the density and velocity fields. In particular, we focus on the role of the shock, condensation fronts, and the flow of the gas transiting between them. The paper is organized as follows. We describe the simulations used in this article in Sect. 2. In Sect. 4 we show and explain the evolution of the 3D magnetic field lines that yield the
alignment of CNM clouds with their local magnetic field. We discuss the implications of our results in Sect. 5. Finally, in Sect. 6, we present our summary and conclusions.
\section{CNM cloud simulations}\label{section_sims_3d}

\subsection{Simulations} \label{sec:sims}

We  performed 2D and 3D numerical simulations of cold atomic clouds formed by the collision of warm atomic gas flowing along the $x$-axis and colliding at the center of the computational domain. The simulations were performed using the adaptive mesh refinement (AMR) code  {\sc{Flash}} version 4.5 \citep{fryxellFLASHAdaptiveMesh2000,2008PhST..132a4046D,DUBEY2009512}, and the ideal magnetohydrodynamics (MHD) multi-wave HLL-type solver \citep{waaganRobustNumericalScheme2011}. Since our goal is to study the alignment of the magnetic field with CNM clouds before self-gravity becomes dominant, neither self-gravity nor any external gravitational potential is included in these simulations. These simulations consider inflow boundary conditions in the $x$ direction and periodic boundary conditions in the other directions.
The initial conditions for both kinds of simulations consist of gas in thermal equilibrium with temperature, $T_{0}=5006.25 \degK$, implying a sound speed $c_{\rm s,0}= 7.36 \ \kms$ for a mean particle mass of $1.27 m_{\rm H}$, and an atomic hydrogen number density, $n_{H,0}=1 \pcc$.

For the 3D simulation, the box size of the simulation is $L= 64 \pc$ and the highest resolution is 0.03125 $\pc$. The gas inside a cylinder of radius $R=16\pc$, length $\ell = 64\pc$, and centered in the middle of the computational domain, has a velocity $u_{0}=\pm14.7 \kms$ along the $x$ direction, with the positive and negative values applying to the left and right of the $x=0$ plane, respectively. We evolve this simulation for 5 Myr.

For the 2D simulation, the box size is $L= 20 \pc$ with a uniform grid and 512 cells per dimension, resulting in a
resolution of 0.039 $\pc$, and we run this simulation for 3 Myr. The main difference is that the collision front in this simulation has a sinusoidal shape, in order to trigger the NTSI, described in \cite{vishniac_1994}. This interface is accomplished by requiring that simulation points with $x < 3.0\sin(\frac{8\pi y }{20}) \pc$ and $x > 3.0\sin(\frac{8\pi y }{20}) \pc$ have a velocity $u_{0}=\pm14.7 \kms$, respectively.
In addition, for the 3D simulation, we add to each velocity component a pseudo-random velocity fluctuation obtained from a Gaussian distribution with zero mean and a standard deviation of $2.85 \kms$. These initial conditions imply an initial Mach number of $M_{\rm s}\approx 2.0$ for both the 2D and 3D simulations.
Furthermore, both simulations consider the multi-phase atomic interstellar medium by including the net cooling function provided in \citet{koyama_2002}\footnote{With the typographical corrections given in \citet{VS_2007}}, where

\begin{equation}
 \Gamma= 2.0 \times 10^{26} \erg \,{\rm s}^{-1}
\label{heating}
\end{equation}
and
\begin{equation}
 \Lambda = \Gamma \Bigr[10^{7} \textrm{exp} \Bigr(\frac{-1\ldotp184\times 10^{5}}{T+1000}\Bigr) +1\ldotp4 \times 10^{-2} \sqrt{T} \textrm{exp} \Bigr( \frac{-92}{T}\Bigr) \Bigr ] \ \textrm{cm}^{3} ,
\label{cooling}
\end{equation}
are the heating rate and the cooling function, respectively.

The initial magnetic field for both simulations is $B_{0}= 3 \muG$ along the $x$-axis, implying an initial Alfv\'en speed of $6.54 \kms$ and an inflow Alfv\'enic Mach number of $M_{\rm A} \approx  \ 2.25$. The magnetization of this simulation results in a plasma $\beta$ of
\begin{equation}
   \beta \equiv \frac{P_{\rm th}}{P_{\rm mag}} = \frac{2 c_{\rm s}^{2}}{u_{\rm A}^{2}} = 2.54.
   \label{beta}
\end{equation}
We note that $\beta$ is formally defined as the ratio of thermal and magnetic pressures, which introduces the factor of 2 in the numerator. However, this additional factor is often omitted in the literature.

Therefore, the initial conditions of these simulations are supersonic, super-Alfv\'enic, and with intermediate magnetization. In Fig. \ref{fig_3d_cd} we show face-on and edge-on column densities of the resulting evolution of the 3D simulation after $5\Myr$. In the following discussion, the highlighted white region located in $x \in [-3.0,3.0] \ \pc$,  $y \in [-1.0,3.0] \  \pc$, and $z \in [2.3,6.3] \ \pc$ will be referred to as R1, whose 3D density structure and magnetic field lines are shown in Fig. \ref{fig_3d_lines},\footnote{This and the other 3D figures were done with the help of Pyvista \citep[][]{pyvista_2019}}
 in which the shock fronts are visible as shaded vertical sheets. We note that the magnetic field lines start to bend at the shock fronts. Additionally, although the magnetic field started nearly horizontally and, thus, perpendicular to the dense layer, at the time shown it has become almost perpendicular to its original orientation ($x$-axis) in some regions; therefore nearly parallel to the dense layer. The bending of the field lines increases over time, as can be seen in the animation accompanying Fig. \ref{fig_3d_lines}, available online.
\begin{figure*}
    \centering
        \includegraphics[scale=0.8]{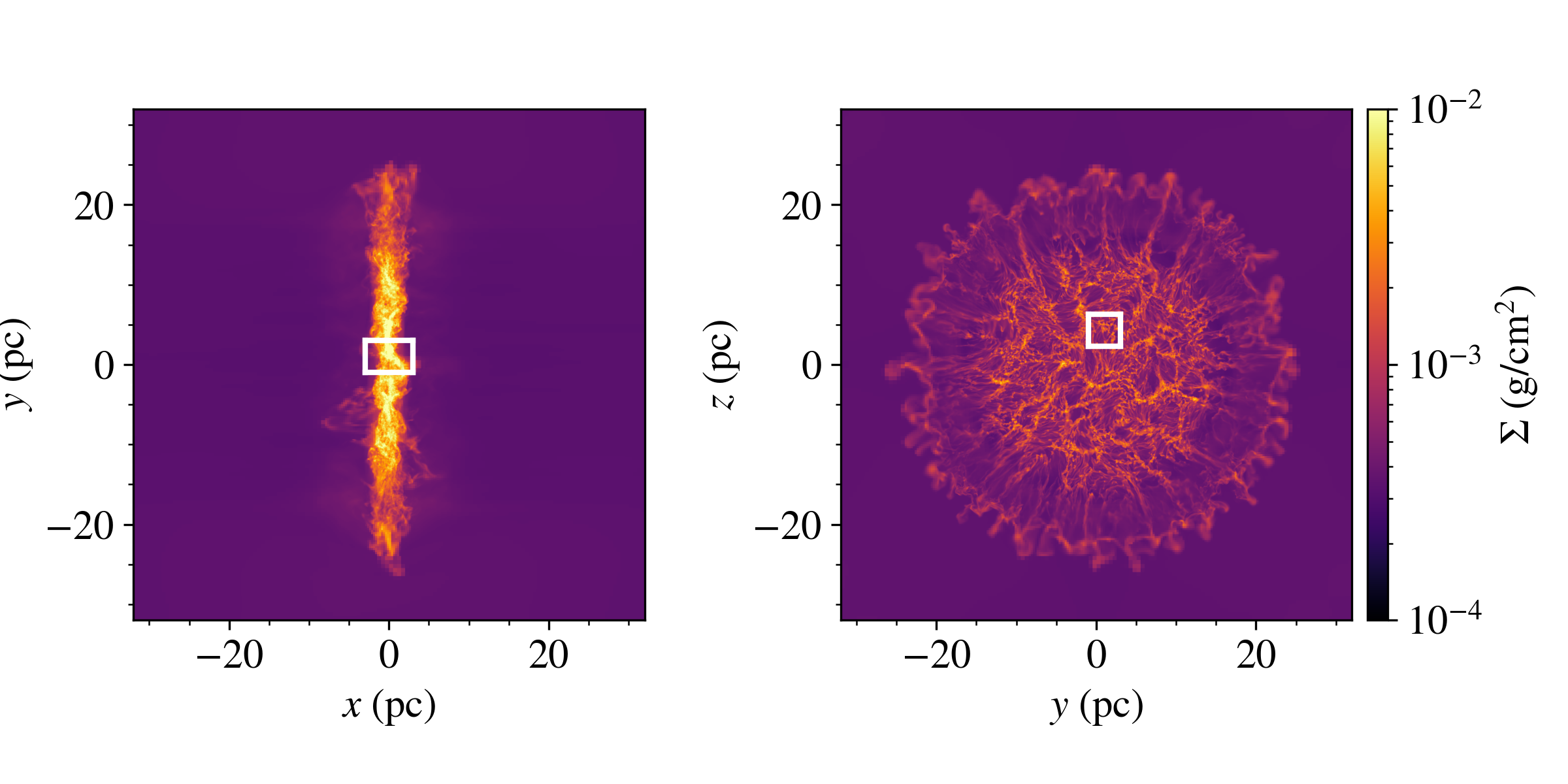}
    \caption{Column density of the CNM cloud's simulation at $5 \Myr$. The highlighted region (R1) is examined in 3D in order to show the magnetic field and
    density morphology.}
    \label{fig_3d_cd}
\end{figure*}
\begin{figure}
\centering
\includegraphics[scale=0.25]{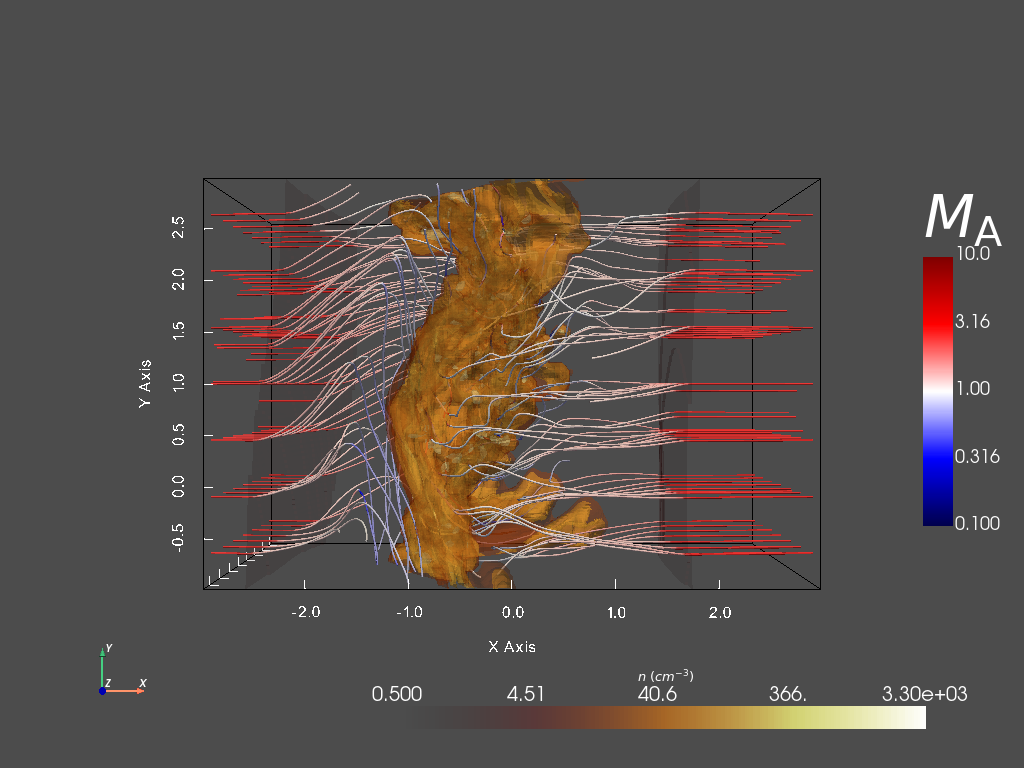}
\caption{Density structure and magnetic field lines of the region R1 are shown after 5 Myr of evolution.  The magnetic field lines are colored by the Alfv\'enic Mach number. We note that at this evolution time, magnetic field lines are almost perpendicular to their original orientation, along the $x$ axis and the change from super-Alfv\'enic to trans-Alfv\'enic of magnetic field lines across the shock. The dark-shaded regions at both sides of the center are the shock fronts where magnetic field lines start to bend. The associated movie is available \href{https://youtu.be/AD86jTnpQzg}{online}.}\label{fig_3d_lines}
\end{figure}

\subsection{The shock-triggered phase transition from WNM to CNM}

The initial condition of these simulations corresponds to atomic WNM gas in thermal equilibrium. The supersonic collision of this gas in the center of the computational domain constitutes a nonlinear perturbation that takes the gas out of thermal equilibrium, to which the gas responds by cooling down toward the CNM phase due to nonlinear TI \citep{Koyama_2000}. As the gas cools and becomes denser behind the shock, the cooling rate increases due to the increasing density, and eventually generates a sharp condensation front, located roughly one cooling length downstream from the shock \citep{VZ_2006}.

In Fig. \ref{fig_pressure_eq}, we show the pressure versus hydrogen atomic number density of region R1 after $5 \Myr$ of evolution. This figure shows that some gas continues in the initial WNM phase in thermal equilibrium since R1 includes pre-shock regions. Most of the gas, however, is not in thermal equilibrium and occupies the region between the shock front and the condensed layer. Another important fraction of the gas is already in the cold phase and back in thermal equilibrium.

\begin{figure}
    \centering
    \includegraphics[scale=0.5]{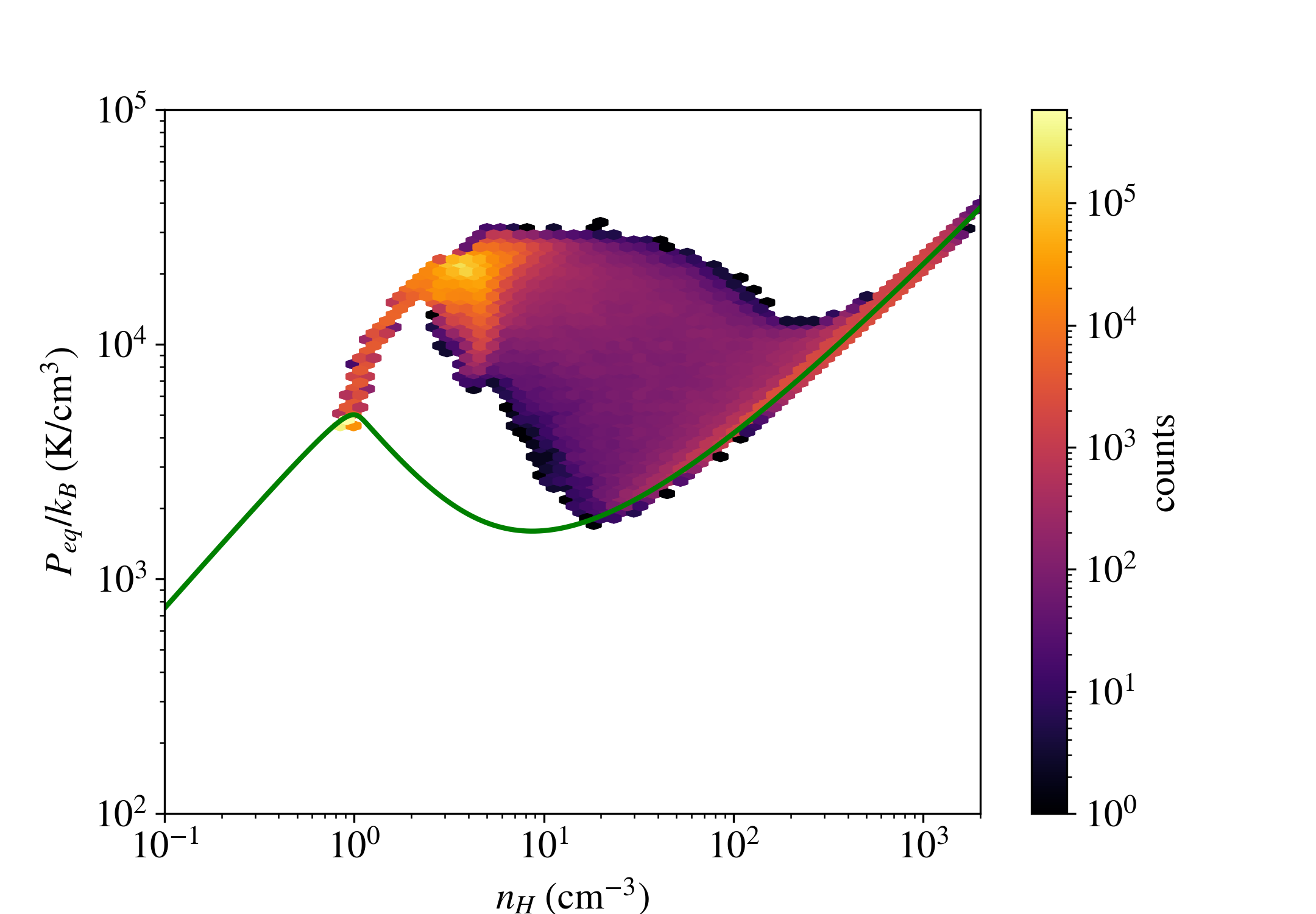}
    \caption{Pressure versus hydrogen number density of the region R1 after 5 Myr of evolution. The green solid line represents the pressure at thermal equilibrium and the color bar represents the counts of the volume-weighted 2D histogram.}
    \label{fig_pressure_eq}
\end{figure}

\section{Alignment of magnetic field lines with cold atomic clouds}

To quantify the relative orientation of magnetic field lines, we used the HRO, which is a statistical tool to measure the angle ($\phi$) between the magnetic field and the density gradient of structures in the ISM, over density intervals\footnote{Note that in this study, we focus in simulations and do not explore the effects of observational effects like
the constraint of only detecting the plane of the sky magnetic field and the relative orientation between the observer and the cloud.} relevant to the multi-phase nature of CNM clouds. These intervals comprise the densities of the post-shock WNM ($n \in [3,10] \ \rm{cm^{-3}}$), the low-density cold neutral gas ($n \in [10,3\times 10^{1}] \ \rm{cm^{-3}}$), the medium-density cold neutral gas ($n \in [3\times 10^{1},10^{2}] \ \rm{cm^{-3}}$), the high-density neutral gas ($n \in [ 10^{2},3\times 10^{2}] \ \rm{cm^{-3}}$), and the density range of the central region ($n \in [ 3 \times 10^{2}, 10^{3}] \ \rm{cm^{-3}}$). We show the resulting HRO in the top panel of Fig. \ref{fig_hro} in terms of $\cos \phi$. Thus, when $\cos \phi = 0$, the magnetic field is parallel to the density isocontours; however, when $\cos \phi = \pm 1$, the magnetic field is perpendicular to the density isocontours. We keep this convention to compare to the 3D HRO diagram presented in \citet{Soler_2013}.  We can see that the HRO for most density intervals peaks at $\cos (\phi) = 0$; in other words, the magnetic field tends to be parallel to the density structures throughout the density range we investigate.

In order to quantify the HRO, \citet{Soler_2013} introduced the shape parameter $\zeta$ defined as
\begin{equation}
  \zeta \equiv \frac{A_{c} - A_{e}}{A_{c}+A_{e}}  ,
 \label{eq_shape}
\end{equation}
where $A_{c}$ is the central area under the HRO diagram located between $\phi \in [75.52^{\circ},104.48^{\circ}]$, corresponding to mostly parallel magnetic field,  while $A_{e}$ is the area under the HRO in the range $\phi \in [0^{\circ},41.41^{\circ}] \cup [138.59^{\circ},180^{\circ}]$, corresponding to mostly perpendicular field. Thus, for a given density interval, if $ 0 < \zeta <1$, the magnetic field lines are parallel to the density gradients, while if $ -1 < \zeta <0$, the magnetic field is perpendicular to the density gradients. Finally, $\zeta$ close to zero happens when there is no clear tendency in the relative orientation.

In the bottom panel of Fig. \ref{fig_hro}, we plot $\zeta$ for the simulation in the density ranges defined above and we consider additional density intervals for the sake of clarity of the trend. This plot shows that the magnetic field becomes increasingly parallel to the density gradient as the density increases up to around $25 \cm^{-3}$ and then the degree of parallel relative orientation decreases.

\begin{figure}
    \centering
    \includegraphics[scale=0.6]{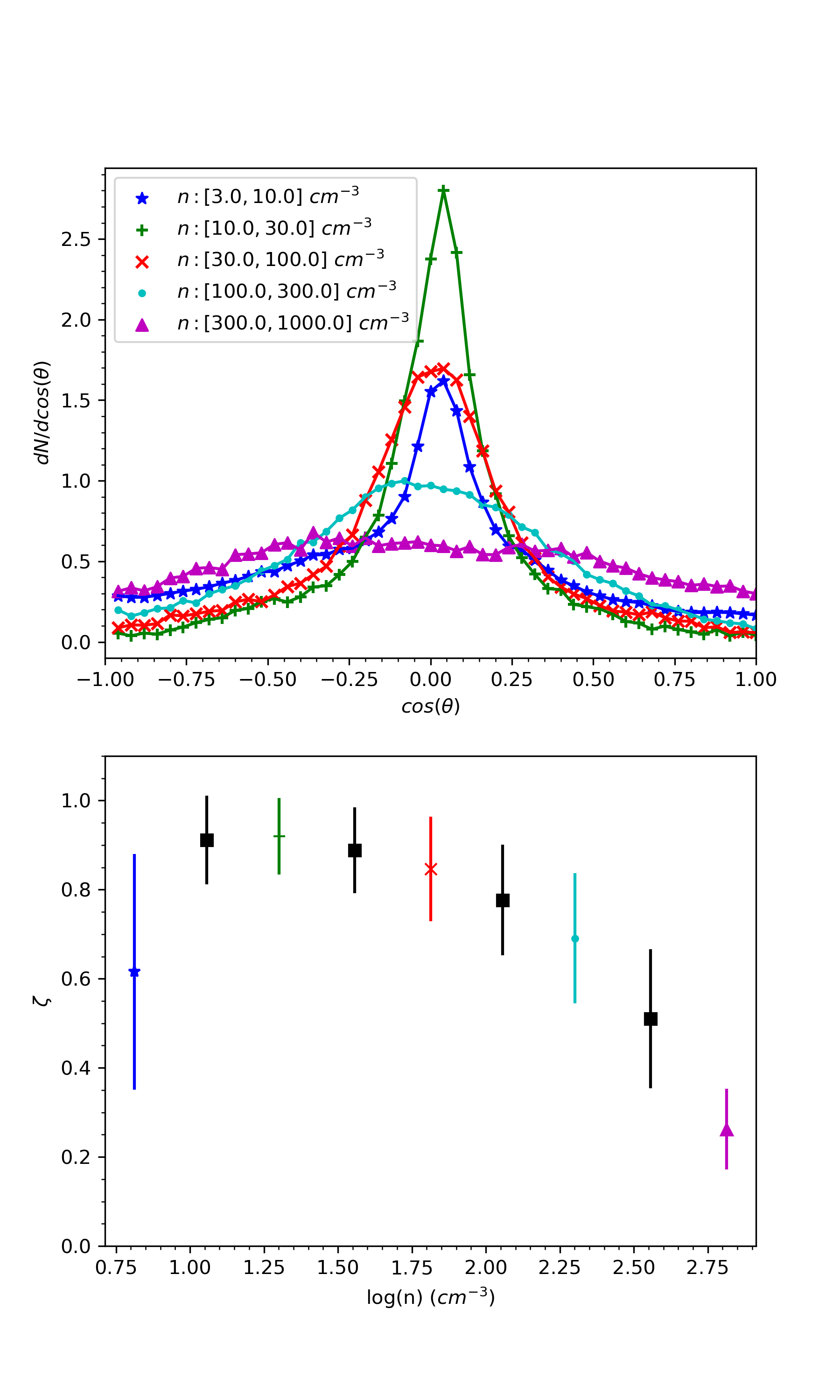}    \caption{Statistics of the relative orientation of the magnetic field and density structures for the 3D simulation. Top: HRO diagram computed at 5 Myr for four different number density intervals. Bottom: Shape parameter (see Eq. 4.3) versus number density, where the error bars were computed using Eq. 5 of \citet{plank_2016_mc_align}. In this panel, shape parameter values in black squares are not shown in the HRO at the top panel. It can be seen that the magnetic field is increasingly parallel over the first three intervals, but then the parallel relative orientation decreases towards higher densities.}    
    \label{fig_hro}
\end{figure}
The trend of the shape parameter shown in Fig. 6 of \citet{Soler_2013} indicates that the relative orientation of the magnetic field and isodensity contours becomes less parallel as the density increases. In that work, the authors sample density values, $n \in [1.6 \times 10^{2}, 3.16 \times 10^{6}] \ \cm^{3}$, in isothermal simulations of cold molecular gas.  In contrast, since we are interested in understanding how this correlation arises as the cloud is assembled, our simulation starts with warm atomic gas leading to density structures with $n \in [6\times 10^{-1}, 2.2\times 10^{3}] \ \cm^{3}$. Therefore, the obtained HRO shape parameter in Fig. \ref{fig_hro} complements the one obtained by  \citet{Soler_2013}, as it corresponds to gas that can be considered the precursor of a molecular cloud. Specifically, the HRO and shape parameter obtained in Fig. \ref{fig_hro} shows an increase in the degree of parallel relative orientation for the first three density intervals, while for the rest of the intervals, we can appreciate a change of this tendency towards a non-preferential relative orientation, which corresponds to the lowest density interval in the results of \citet{Soler_2013}.

In Fig. \ref{fig_hro_int_3d}, we show the density structures
between the four highest density intervals used to obtain the HRO and shape parameter of Fig. \ref{fig_hro}. It can be seen from this figure that the magnetic field (black arrows) tends to be parallel to the density structures for the four density intervals. However, for the highest density interval, the magnetic field does not follow this general trend in some regions, leading to the observed change in the shape parameter.
\begin{figure*}
   \centering
   \includegraphics[scale=0.4]{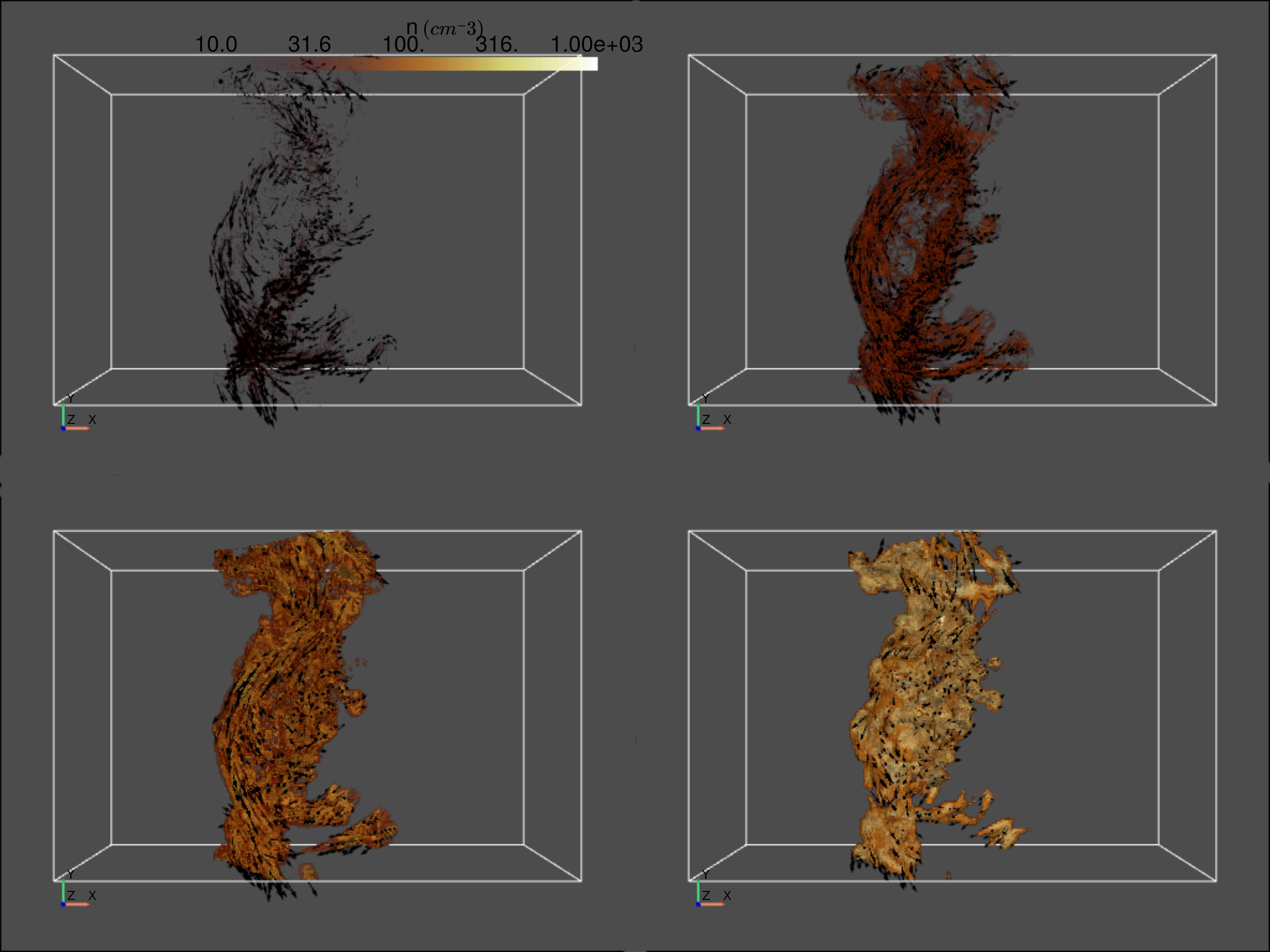}
\caption{Regions used to compute the HRO diagram and the shape parameter shown in Fig. \ref{fig_hro}. The yellow surfaces are density isocountours and the dark arrows represent the magnetic field. The top-left panel corresponds to gas with density $n \in [10,30]\ \cm^{-3}$,  top-right panel to $n \in [30,100] \ \cm^{-3}$,  bottom-left to $n \in [100,300] \ \cm^{-3}$, and  bottom-right to $n \in [300,1000] \ \cm^{-3}$. It can be seen that for the four density intervals shown here, the magnetic field is noticeably parallel to the density structures. However, the magnetic field shows a deviation from this trend for the highest density interval in the bottom right panel. The associated movie is available \href{https://youtu.be/ySIy-ZfnutE}{online}.}
\label{fig_hro_int_3d}
\end{figure*}

\section{Magnetic field line evolution}\label{section_lines_evol}

To understand the alignment of the magnetic field with CNM clouds shown in the previous section, we followed the evolution of the magnetic field lines. The resulting configuration of the
3D density structures and magnetic field lines of region R1 after 5 Myr of evolution is shown in Fig. \ref{fig_3d_lines}, where we can see the shock fronts at each side of the central condensation region. The magnetic field lines start bending at these shock fronts, along their way from the center of the computational domain.

As can be seen in the provided animation (Fig.\ \ref{fig_3d_lines}), the magnetic field lines change their direction from being nearly parallel to the $x$-axis at early times to being mostly perpendicular to it after $5 \Myr$ in the neighborhood of the dense layer. We investigate  this process below.

\subsection{Magnetic field amplification by MHD shocks}\label{subsec_shock}

 As seen in Fig. \ref{fig_3d_lines}, and considering the shock front located at the right of the condensation layer, we see that the angle between the upstream magnetic field and the normal to the shock front satisfies $\theta \approx 0$ for all the magnetic lines shown. The small variations of $\theta$ around zero are due to the fact that the shock front is not a plane when it moves away from the central region because of the fluctuations added to the inflow velocity in the simulation setup.

Following \citet{delmont_keppens_2011}, the fast magnetosonic speed $u_f$ is defined as:
\begin{equation}
 \centering
 u_{f}^2= \frac{1}{2} \left(c_{s}^2 +u_{\rm{A}}^2 + \sqrt{(c_s^{2}+u_{\rm{A}}^{2})^{2}-4 u_{\rm{A,n}}^{2}c_{s}^{2}}\right),
 \label{eq_fast_speed}
\end{equation}
where $u_A$ is the Alfv\'en speed and $u_{A,n}$ is its component normal to the shock front. Defining $u_{n}$ as the flow speed normal to the shock, the flow is referred to as \textit{superfast} when $|u_{n}|>u_{f}$. Since $u_{A,n} \approx u_{A}$ and $u_{n} \approx u_{0}$ for the preshock flow in the 3D simulation described in Sect. \ref{section_sims_3d}, it can be seen from Eq. \eqref{eq_fast_speed} that this flow is superfast. The downstream flow, just after the shock front, becomes trans-Alfv\'enic, as we can see from Fig. \ref{fig_3d_lines}. Therefore, the relation
$u_{f} > |u_{n}| > u_{A,n}$, which characterizes a subfast flow, is satisfied downstream.

This type of MHD shock, which goes from a superfast to a subfast
flow, is called a fast MHD shock \citep[]{delmont_keppens_2011}. Its main feature
is its refraction of the magnetic field away from the shock normal, due to the amplification of the magnetic field component parallel to the shock front. This amplification is given by
\begin{equation}
 %\label{eq_mhd_shock}
 B_{\parallel, 2}=\frac{r_{\rho} B_{\parallel,1}(M_{A,1}^2-\cos^{2} \theta)}{M_{A,1}^2-r_{\rho}\cos^{2} \theta},
 \label{eq:B_amplif}
\end{equation}
where $B_{\parallel,1}=B_{1}\sin \theta$ and $B_{\parallel,2}$ are the upstream and downstream magnetic field components parallel to the shock front, $r_{\rho}=\rho_{2}/\rho_{1}$ is the ratio of the downstream density ($\rho_{2}$) to the upstream ($\rho_{1}$) density, $M_{A,1}$ is the Alfv\'enic Mach number of the upstream gas, and $\theta$ is the angle between the vector normal to the front-shock and the upstream magnetic field $\bm{B}_1$.
Since this amplification depends on the angle $\theta$, the fluctuating curvature of the shock front at different positions yields the inhomogeneous downstream magnetic field pattern when the shock front travels away from the central region, early on  in the evolution timescales (see Fig. \ref{fig_3d_lines}).

\subsection{Line bending analysis}\label{subsection_line_model}
To understand how magnetic field lines change their original direction in the post-shock region, we consider the induction equation in ideal MHD,
\begin{equation}
 \frac{\partial \bm{B}}{\partial t} = - \bm{B} \nabla \cdot \bm{u} - (\bm{u}\cdot \nabla)\bm{B} + (\bm{B} \cdot \nabla )\bm{u}.
 \label{eq_flux_f2}
\end{equation}

\subsubsection{Line bending by a compressive flow}\label{line_compressive}
The following analysis  comes after the magnetic field component parallel to the shock front is amplified
by the fast MHD shock, namely, in a region containing cooling and thermally unstable gas. After the amplification, magnetic field lines adopt the shape represented in Fig. \ref{fig_line_bend}, where the magnetic field component parallel to the shock front, $B_{y}$, has been amplified and the magnetic field component perpendicular to
the shock front, $B_{x}$, remains constant. This is in agreement with the jump condition for the magnetic field.

Furthermore, we assume that the flow speed decreases along $x$  in the post-shock region; namely, $u_{x}=u_{x}(x)$ and $\partial u_{x}/\partial x < 0$, which represents the compression caused by the cooling of the gas as it travels downstream. Finally, we disregard the downstream component of the velocity parallel to the shock front, $u_{y}$ and $u_{z}$, to analyze the effect of the compression alone.
\begin{figure}
    \centering
        \includegraphics[width=0.95\columnwidth]{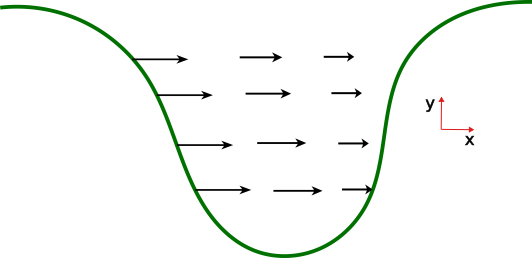}
    \caption{Initial line bending model. The magnetic field is represented by the green line, while the velocity field is represented by the black arrows.}
        \label{fig_line_bend}
\end{figure}
To validate these assumptions, in the left panels of Fig. \ref{fig_ray}, we plot the relevant physical quantities at a time of $t = 0.7 \Myr$, along a ray parallel to the $x$ axis passing through a region where this amplification becomes higher at later evolutionary times. In the right panels of this figure, we show the same quantities at the later time $t = 5$ Myr.

The shock fronts are seen as the sharp jumps at $x \approx -1.5$ pc and $x \approx +1.5$ pc in the gas density in the top-left panel and in the $u_x$ velocity component in the middle-left panel. Also, the condensed region is seen as the sharp peak at $x = 0$ in the gas density profile. The middle left panels also show that in addition to the discontinuity at the shocks, the inflow velocity, $u_{x}$, smoothly decreases in the shock-bounded region, in sync with the density increase, indicating a compressive velocity gradient along the $x$-direction in this region. The compressive gradient is even steeper in the condensed layer. Also, $u_{y}$ and $u_{z}$ remain negligible in the shocked region, supporting our assumptions. Finally, in the bottom-left panel, we see that the fluctuation of $B_x$ remains within $\lesssim 20\%$ of its mean value, so it is negligible to the first order. On the other hand, we also see that the $B_y$ and $B_z$ components show noticeable fluctuations immediately downstream from the position of the shocks, indicating the refraction of the field lines at the shocks.
\begin{figure*}
 \centering
 \includegraphics[scale=0.8]{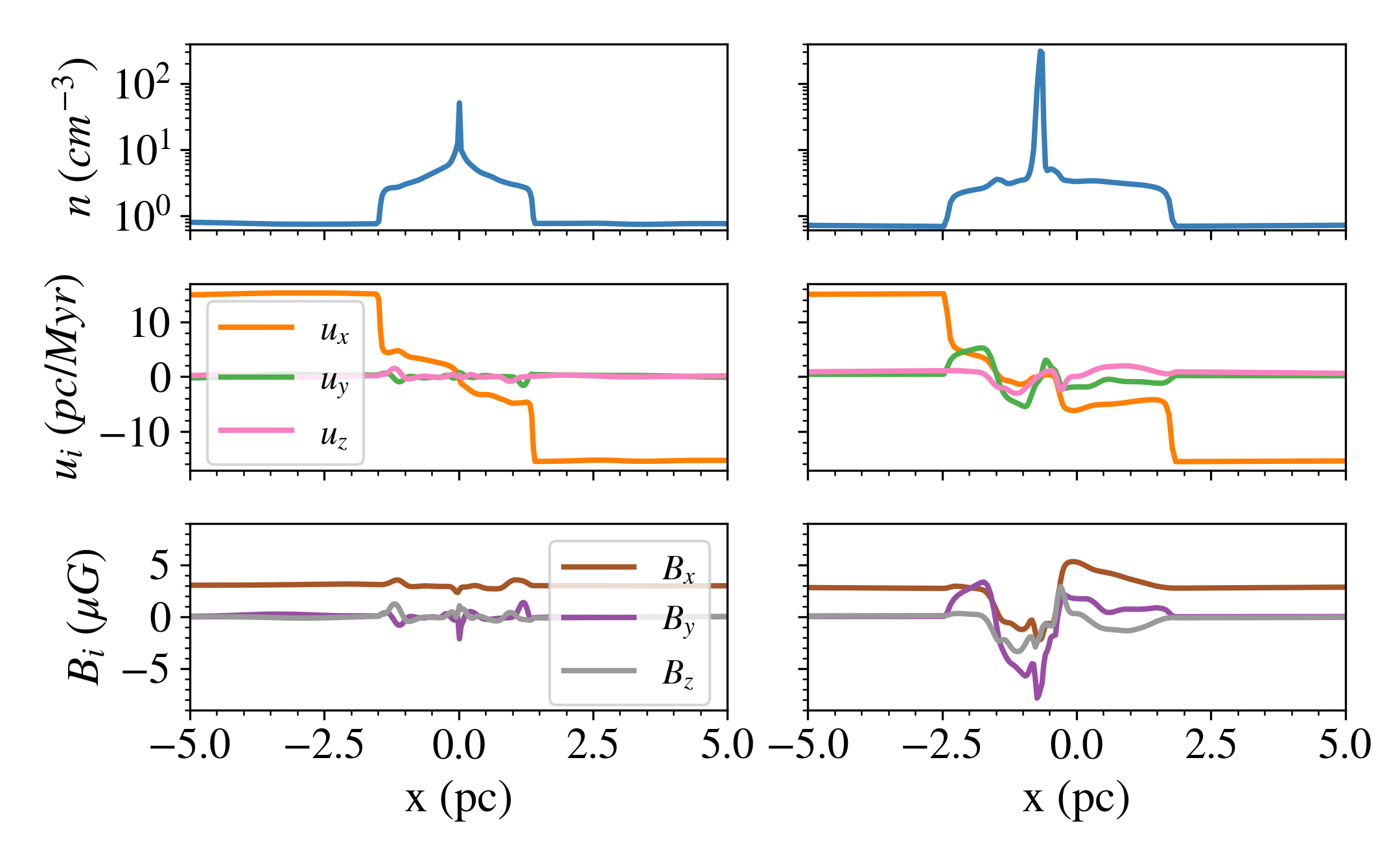}
 \caption{Number density (top), velocities (middle), and magnetic field components (bottom) along a ray parallel to the $x$ axis at 0.7 Myr (left panels) and 5.0 Myr (right panels) of evolution. The compressive velocity field, $u_{x}$, and the magnetic seed amplifications at the MHD fast shock are given in the middle-left and bottom-left panels, respectively, and the amplified $B_{y}$ and $B_{z}$ components in the bottom-right panel.}
 \label{fig_ray}
\end{figure*}
Therefore, the assumptions $B_{y}=B_{y}(x)$, $B_{x}=C$, where $C$ is a constant,  $u_{x}=u_{x}(x)$, with $\partial u_{x} / \partial x < 0$, $u_{y}, u_{z} \rightarrow 0$, and solving for the $B_{y}$ component, reduce Eq. \eqref{eq_flux_f2} to
\begin{equation}
 \centering
 \frac{\partial B_{y}}{\partial t} = -B_{y}\frac{\partial u_{x}}{\partial x} -
 u_{x}\frac{\partial B_{y}}{\partial x}.
 \label{eq_line_model_1}
\end{equation}
This equation can also be written in Lagrangian form as
\begin{equation}
 \centering
 \frac{d B_{y}}{d t} =  - B_{y} \frac{\partial u_{x}}{\partial x}.
 \label{eq_line_model_2}
\end{equation}
Therefore, since $\partial u_{x} / \partial x < 0$, Eq.\ \eqref{eq_line_model_2} implies that $d B_y/dt$ has the same sign as $B_y$; thus, the magnetic field component $B_{y}$ %\footnotemark
is always amplified by the downstream compressive velocity gradient. This amplification results in the magnetic field aligning to the condensation plane where CNM clouds form. The amplification of the magnetic field components parallel to the dense layer ($B_y$ and $B_z$ in the simulation) is clearly seen in the bottom right panel of Fig. \ref{fig_ray}, which shows the various physical quantities at 5 Myr. Specifically, it is apparent that these field components have grown in the shock-bounded region to values comparable to the initial value of $B_x$, indicating their amplification by the compressive velocity field.

\subsubsection{Line bending at curved interfaces}

Another possible mechanism for aligning the magnetic field with the density structures occurs when the collision interface is curved, rather than flat, as, for example, in the case of the NTSI \citep{vishniac_1994}. To investigate this mechanism, we also ran 2D simulations with the same initial conditions and physics according to the 3D simulation described in Sect. \ref{section_sims_3d}, but with a curved collision interface, obtained by adding a sinusoidal displacement perturbation (a ``bending mode'' perturbation). In the left panel of Fig. \ref{fig_ntsi_curv}, we show a very early stage of this simulation. In this case, the obliqueness of the interface implies the existence of a component of the incoming flow tangential to it,
while the perpendicular component is reduced across the shock. This causes the flow to change direction at the interface,  now that is oblique to the original magnetic field direction. As it is trans-Alfv\'enic, this oblique post-shock flow can begin bending the field lines. The situation is symmetric on both sides of the layer, thus generating a shearing velocity field with opposite directions at opposite sides due to the different concavity of the collision interface. This generates an ``S'' shape of the magnetic field lines across the shocked layer. A later stage of this simulation is shown in the right panel of Fig. \ref{fig_ntsi_curv}, showing that the flow tends to be sub-Alfv\'enic in the condensed regions.
\begin{figure*}
 \centering
 \includegraphics[scale=0.8]{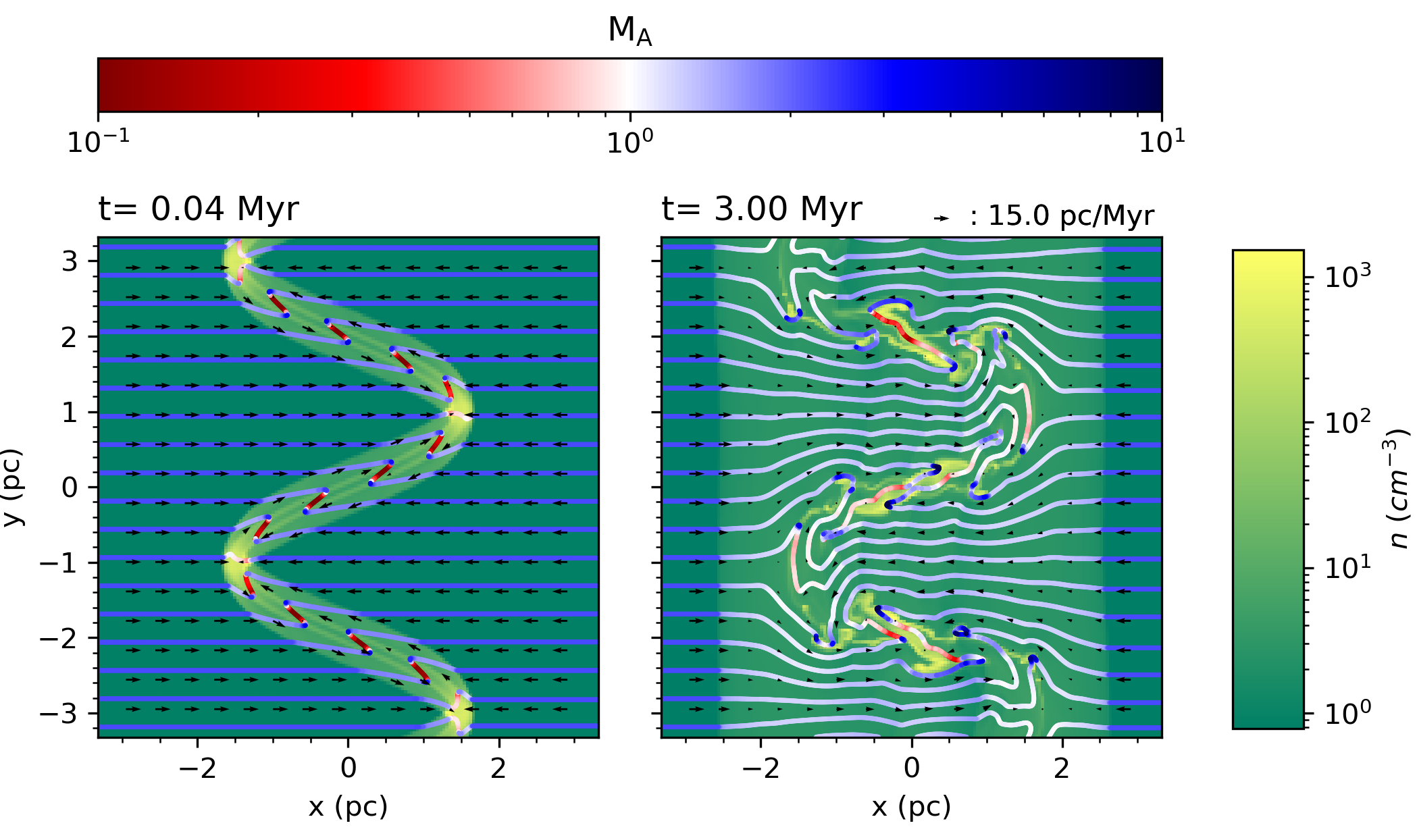}
 \caption{2D simulation of warn atomic colliding flows with a curvilinear collision interface at 0.04 Myr (left panel) and 3.0 Myr (right panel). The two
 color bars represent the number density and the Alfv\'enic Mach number.}
 \label{fig_ntsi_curv}
\end{figure*}
For this collision interface, the line-bending analysis differs from that described in Sect. \ref{subsection_line_model}. In this case, we will have a velocity field like the one represented in Fig. \ref{fig_line_bend_ntsi}, where the left panel represents an unperturbed downstream magnetic field line and the right panel represents a perturbed one. In the left panel, we consider a local system of coordinates centered at the point where the $u_{y}$ is maximum. The magnetic field line is represented in green, the $x$-axis is parallel to the field line, the $y$-axis is perpendicular to it, and the velocity field is represented by black arrows.
\begin{figure}
\centering
 \includegraphics[scale=0.33]{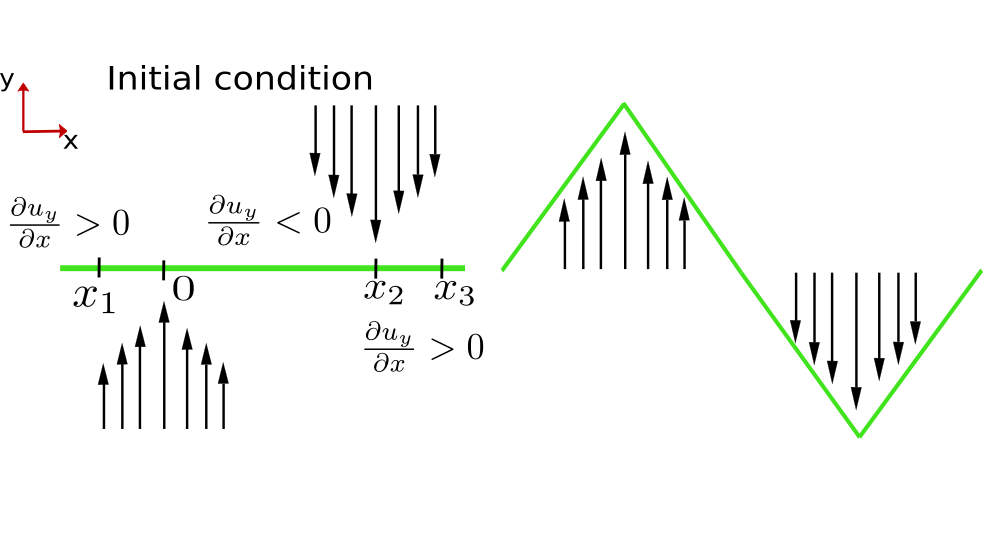}
 \caption{Line bending model for the colliding flows with curvilinear collision interface. The left panel represents
 the initial state and the right one the evolution according to the analytical model given by Eq. \eqref{eq_line_model}. In this sketch, the black arrows are the velocity field.}
 \label{fig_line_bend_ntsi}
\end{figure}
Thus, initially, $B_{y}=0$ and $B_{x} = C$, where $C$ is a constant. Then, considering $u_{x}$, $u_{z}  \rightarrow 0$ and $u_{y}= u_{y}(x)$   from Eq. \eqref{eq_flux_f2}, we obtain:%
\begin{equation}
   \label{eq_line_model}
   \frac{\partial{B_{y}}}{\partial t} = B_{x}\frac{\partial u_{y}}{\partial x} .
\end{equation}
We can then consider the previous equation in three different regions in the left panel of Fig. \ref{fig_line_bend_ntsi}. First, in the region $x \in [x_{1}, 0]$, we have $ \partial u_y/\partial x \ > \ 0$, and so $\partial B_y/\partial t \ > \ 0$. Second, in the region $x \in [0, x_{2}]$, we see that $\partial u_y/\partial x \ < \ 0$ and, thus, $\partial B_y/\partial t \ < \ 0$. Finally, in the region $x \in [x_{2},x_{3}]$, $ \partial u_y/\partial x \ > \ 0$, thus $\partial B_y/\partial t \ > \ 0$.
Therefore, for these three regions, the sign of $\partial B_y/\partial t$ explains the deformation of the magnetic field line having an "S" shape morphology, where each convex part
aligns the direction of the flow (see Fig. \ref{fig_ntsi_curv}, left panel,  and Fig. \ref{fig_line_bend_ntsi}).

\section{Discussion}\label{section_discussions}

\subsection{The role of the pre-condensation shock and the velocity gradient in the alignment of the magnetic field}\label{shock_role}

As we see from the 3D simulation, magnetic field lines change their orientation at the fast MHD shock front, being refracted away from the shock normal. This results in a fluctuating orientation of the field lines behind the shock. These fluctuations are then further amplified by the compression generated by the downstream decelerating gas. The combined action of the fast MHD shock and the velocity gradient can be thus described as the shock producing the seeds of the magnetic field fluctuations, along with the compressive velocity gradient amplifying them.

It is important to note that in this work, we have not varied the relative orientation between the upstream magnetic field and the shock front
of the system in the initial condition. However, there is a small range of angles between them due to the departure of the shock front from a perfectly flat plane due to the small fluctuations added to the inflow velocity. The role of the initial angle between the magnetic field and the shock front has been studied in \citet{Inoue_2016}, who found that the number of filamentary CNM clouds or fibers oriented perpendicular to the magnetic field increases with the angle between the upstream magnetic field and the shock front in simulations without an initial velocity dispersion. However, when an initial velocity dispersion was included, those authors found that fibers tend to be oriented in the direction of the local magnetic field. For this reason, they concluded that the formation mechanism of fibers and their alignment with the local magnetic field is the turbulent shear strain, which was also identified as the reason for the elongation of CNM clouds to form filaments by \citet{Hennebelle_2013}.

It is also important to mention that the role of MHD shocks and the nature of the velocity gradient in the region between the shock and the condensation fronts in the evolution of magnetic field lines has not been explored before.
Previous studies have focused, for example, on whether the magnetic field lines get reoriented along the flow or vice versa, depending on the inflow Mach number \citep[e.g.,] [] {Hennebelle_2000}, as well as on whether the filaments are produced by shocks or by stretching due to the turbulent strain \citep[e.g.,] [] {Hennebelle_2013} and its orientation with respect to the magnetic field direction \citep[e.g.,] [] {Inoue_2016}. However, to our knowledge, the specific combined action of the shock front and the deceleration due to the condensation has not been described before. This result implies that supersonic and super-Alfv\'enic compressions in the atomic medium, even when exerted along the magnetic direction, generally will not preserve the original magnetic field direction.

\subsection{The role of the velocity gradient in the relative orientation of the magnetic field with substructure in MCs.} \label{sec:vel_grad}

\citet{Soler_2017} proposed that the relative orientations between the magnetic field and density structures $\phi= 90^{\circ}$ and $\phi=0^{\circ}$ might be equilibrium points. However, the specific physical mechanism behind their result remains unknown. In this work, we have proposed that it is the combined action of the fast MHD shock and the compressive velocity, resulting from the gas settlement onto the dense layer. This leads to the alignment of the magnetic field with the dense layer as a consequence of the amplification of the magnetic field components perpendicular to the shock front.

Since we have focused on non-gravitational CNM clouds, we have not numerically explored how $\phi$ switches from $\sim 90^\circ$ to $\sim 0^{\circ}$. However, a discussion similar to that in Sect. \ref{line_compressive} leads us to speculate that the $\phi \sim 0^{\circ}$ configuration may arise in the presence of a stretching velocity field; indeed, this would be the case of the tidal flow into the gravitational well of a strongly self-gravitating cloud. In this case, letting $x$ again be the direction of the flow and $B_{y}$ the magnetic field perturbation perpendicular to that direction, Eq. \eqref{eq_line_model_2} for a positive velocity gradient implies that $d B_y/dt$ has the opposite sign to $B_y$, straightening the field lines. Thus, we suggest that the induction equation in the presence of a compressive or stretching velocity field leads to the $\phi = 90^{\circ},0^\circ$ equilibrium configurations found by \citet{Soler_2017}, respectively, justifying their speculation that these values may be attractors. This also suggests a mechanism for the parallel relative orientation of the magnetic field to non-self-gravitating structures and its perpendicular relative orientation to self-gravitating ones.

\subsection{The effect of strong cooling on the development of the NTSI}

Regarding the development of the NTSI, \citet{vishniac_1994} found that the requirement for this instability to grow is that the displacement of the cold slab is larger than its thickness. This condition can only be satisfied when there is a high compression ratio across the shock, yielding a very thin shocked layer. In the isothermal case, this requires that $M_{\rm s}^{2} \gg 1$. The 3D simulation described in Sect. \ref{section_sims_3d} gives us $M^{2}_{\rm s}=4.0$, which is not too high. However, our simulations include strong cooling leading to TI, which produces a much stronger compression of the condensed layer and a much thinner slab dimension, even for moderate Mach numbers \citep{VS+96}. Thus, it is not difficult to fulfill the requirement for the development of the NTSI at the condensed, rather than the shocked, layer \citep{Hueckstaedt_2003}. This is demonstrated by the growth of the bending mode perturbation (the increase in the curvature) for the dense layer in our 2D simulation. This means that NTSI can be triggered by not-so-strong shocks in the strongly cooling case.

\subsection{The effect of the magnetic field on the NTSI and the shear strain} \label{sec:B_on_NTSI}

The NTSI is one of the possible mechanisms yielding the shear strain proposed by \citet{Hennebelle_2013} to be responsible for the elongation of filamentary CNM clouds and the alignment of these structures with the local magnetic field, since it produces the momentum transport from the original inflow direction to that parallel to the dense layer in the regions around the nodes, as can be seen in Fig. \ref{fig_ntsi_curv}. However, \citet{Heitsch_2007} found that when the magnetic field is aligned with the inflow, it tends to weaken or even suppress the NTSI due to the magnetic tension counteracting the transverse momentum transport. On the other hand, the NTSI can contribute to the change of direction of the magnetic field if the flow surrounding the dense layer remains at least trans-Alfv\'enic, so that it has sufficient energy to bend the field. This condition is indeed satisfied by the flow between the shock and the condensation front, as can be seen for the 2D simulation in both panels of Fig. \ref{fig_ntsi_curv}. We note, incidentally, that the flow inside the dense layer is generally sub-Alfv\'enic.

Another source of shear strain that does not require the NTSI is observed in our 3D simulation, which does not include an initial bending-mode perturbation to the locus of the collision front. This arises later in the evolution, when the magnetic field lines have already been dragged and bent by the compressive velocity field. At this time, the trans-Alfv\'enic condition of the post-shock flow allows for the magnetic field to partially re-orient the gas flow along them. Since the field lines have been oriented nearly parallel to the dense layer by the compressive post-shock flow, the velocity field is also oriented in a similar way and in opposite directions on each side of the dense layer; therefore, this adds a strong shear component to the flow around the layer. We can see one example of this situation in Fig. \ref{fig_magnetic_shear}, where the velocity field is represented by dark arrows and the magnetic field lines are color coded with the Alfv\'enic Mach number. In this figure, the troughs and peaks of the CNM clouds do not show corresponding converging and diverging velocity fields, as they would correspond to the NTSI.  Thus, the structure at this point does not appear to have been formed by this instability. %(See Figure \ref{fig_ntsi}).

\begin{figure*}
\centering
\includegraphics[scale=0.4]{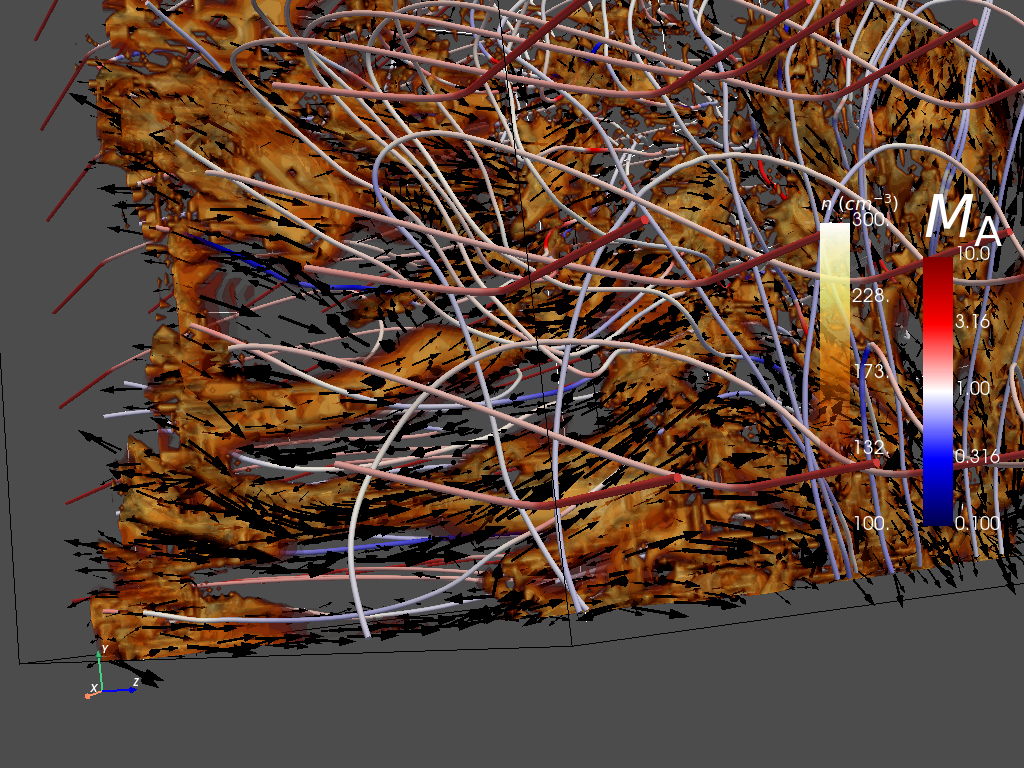}
\caption{Couple of filamentary CNM clouds located in the bottom-left corner. This 3D visualization also corresponds to R1 after 5 Myr of evolution. In the provided animation
showing this region, it can be seen a turbulent shear strain velocity field along their main axis and aligned with the local magnetic field. Note: the velocity field on the filamentary CNM clouds is different from the one expected in the case of the NTSI. The associated movie is available \href{https://youtu.be/saajb-4k7IU}{online}.}
\label{fig_magnetic_shear}
\end{figure*}

\subsection{Magnetic inhibition of turbulence generation} \label{sec:turb_suppr}

It has been noted in previous works that MHD simulations of cloud formation are less turbulent and show more filamentary structure than pure hydrodynamical simulations \citep[e.g.,][]{ Heitsch_2007,Heitsch_2009, Hennebelle_2013,zamora-aviles_magnetic_2018}. The generation of turbulence in curved compressed layers is due to the KHI instability, which, in turn, is triggered by the shear flow produced by the NTSI \citep[e.g.,][]{vishniac_1994, blondin_1996,heitsch_birth_2006}. Therefore, the magnetic tension, which opposes the vorticity generation by the shear flow across the dense layer,
%a magnetic field aligned with the shear flow generated by the NTSI
may suppress the development of the KHI and, as a consequence, the generation of turbulence.
Indeed, a 2D numerical simulation without the magnetic field exhibits a much stronger turbulence level, as shown in Fig. \ref{fig_turbulence}.

\begin{figure}
    \centering
    \includegraphics[scale=0.6]{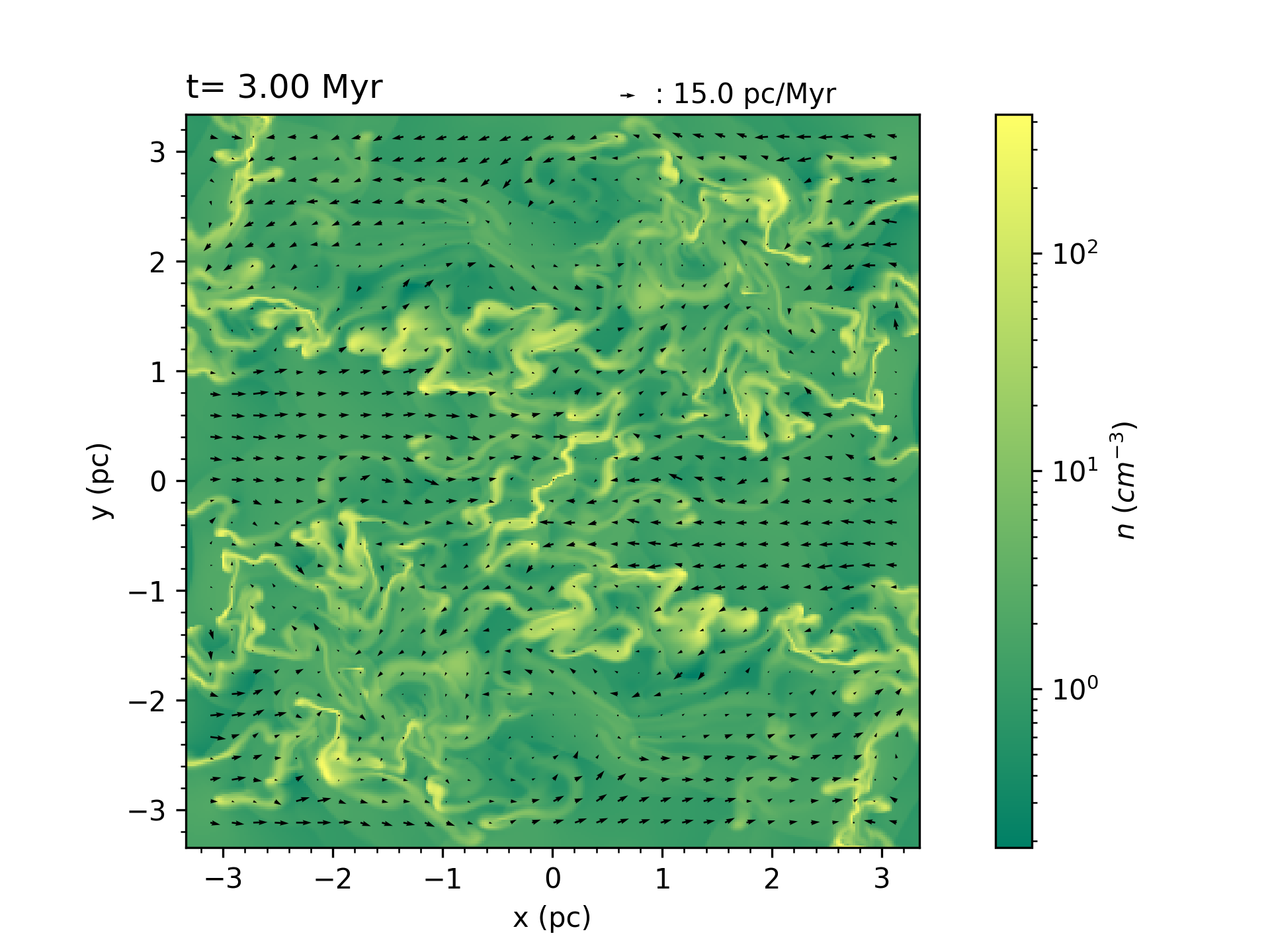}
    \caption{Non-magnetic version of the 2D simulation shown in Fig. \ref{fig_ntsi_curv}.}
    \label{fig_turbulence}
\end{figure}

\subsection{Comparisons with previous work} \label{sec:comparison}

Our result shows that the super-Alfv\'enic nature of the initial inflow in our simulations (and its continuation downstream as a trans-Alfv\'enic flow) allows for the dragging and amplification of the magnetic field. This is in agreement with \citet{skalidis_2022}, whose observations reported trans-Alfv\'enic turbulence in an {\sc Hi}-H$_2$ transition region. According to these authors, atomic gas might accumulate along magnetic field lines, which is also in agreement with our results (see Fig. \ref{fig_magnetic_shear}).

In this work, our simulations consider only the formation of clouds by the collision of converging warm atomic flows. However, the main physical processes responsible for the alignment of magnetic field lines and density structures, fast MHD shocks, converging flows, and the NTSI, can also be present at the interfaces between interacting wind-blown bubbles and/or supernova shells. Therefore, the MHD shocks and the NTSI could also stand as important physical mechanisms behind the magnetic field alignment with fibers found in these types of objects by \citet{clark_2014}. 

\cite{ntormousi_2017} studied the role of magnetic fields in the structure and interaction of supershells without finding a preferred relative orientation between CNM clouds and their local magnetic field, in apparent contradiction with our results and a number of corroborating observations. However, upon closer inspection, this can be understood because their simulations have physical conditions that do not favor the development of the physical mechanisms responsible for the alignment reported in this work. For example, the colliding flows are subsonic when they meet at the center of the computational domain. This implies that there are no shocks generated in this region and the compression of the gas is weaker, thus preventing the onset of the mechanism we have identified in this work. Furthermore, the weaker compression could also lower the probability that the NTSI would end up triggered.

\section{Summary and conclusions}\label{section_conclusions}

In this work, we have investigated the physical mechanisms that might be responsible for the observed alignment of magnetic field and gravitationally unbound CNM clouds formed by the collision of converging warm atomic gas streams. To this end, we tracked the evolution of magnetic field lines caused by moderately supersonic compression in a 3D simulation characterized by typical conditions of the warm ISM. We found that the lines become perpendicular to their original orientation and end up aligned with the forming density structures. 

The alignment occurs because the compression produces a fast MHD shock at each side of the collision plane, along with a condensation front, which delimits the dense layer, roughly one cooling length behind the shock. The shock amplifies the magnetic field components parallel to the shock front, causing the refraction of the field. In addition, the shock heats the gas adiabatically and takes it out of thermal equilibrium. As the flow downstream from the shock cools down and approaches the condensation front, it decelerates and becomes denser, developing a compressive velocity gradient that further amplifies the field components parallel to the shock front, eventually leaving the field lines nearly parallel to the surface of the dense layer. Moreover, the bent magnetic field lines reorient the flow and generate a shear flow around the dense layer.

We also provide an analysis of the induction equation to understand the amplification of the magnetic field components parallel to the fronts by the compressive velocity gradient (Eq.\ [\ref{eq_line_model_2}]). This analysis shows that the time derivative of a field component parallel to the fronts and the longitudinal velocity gradient have opposite signs. Therefore, this magnetic component gets amplified when the velocity gradient is compressive, changing the orientation of the field to aligned with the compressed layer. This implies that moderately supersonic compressions in the WNM (even when initially aligned with the magnetic field) generate a strong field component perpendicular to the initial flow direction, so that the field becomes transverse to the initial compression direction. This extends the results of \citet{Hennebelle_2000} on the field bending via oblique flows to the initially parallel compression case.

From the same equation, we conclude that a stretching velocity gradient should cause a reduction of the parallel component, leading to a straightening of the field lines, thus orienting them perpendicular to the density structures. We speculate that this is the mechanism occurring during the growth of self-gravitating structures, where the flow accelerates inwards, thus producing a tidal stretching velocity pattern. This also provides a possible explanation for the perpendicular relative orientation of the field lines around self-gravitating molecular cloud filaments.

In conclusion, we have found that a settling (i.e., decelerating) flow, occurring as a result of the condensation of the gas by TI orienting the lines parallel to the density structures; meanwhile a stretching (accelerating) one, such as infall into a potential well, orients the field lines perpendicular to the density structures. This could may comprise the physical mechanism behind the stationarity of these configurations found by \citet{Soler_2017}.
Finally, we also found that under typical conditions of the ISM, the flow upstream from the shock front is super-Alfv\'enic and becomes trans-Alfv\'enic downstream. This allows the velocity field to bend and drag the magnetic field lines.

Our study extends previous studies of the formation of CNM clouds and the relative orientation of the magnetic field with the density structures. In particular, we have identified the crucial role played by the pre-condensation shock and the nature (compressive or expansive) of the velocity gradient in generating a parallel or perpendicular relative orientation, respectively.

\begin{acknowledgements}
We are grateful to Susan Clark and Laura Fissel for useful comments and suggestions and to an anonymous referee, whose comments helped us in improving the manuscript.
This research was supported by a CONAHCYT scholarship.
GCG and EVS acknowledge support from UNAM-PAPIIT grants IN103822 and IG100223, respectively. GGM acknowledges partial financial support from ANID/ Fondo 2022 ALMA/31220028. In addition, we acknowledge Interstellar Institute's program “With Two Eyes” and the Paris-Saclay University's Institut Pascal for hosting discussions that nourished the development of the ideas behind this work.
\end{acknowledgements}

\end{nolinenumbers}
\bibliographystyle{./bibtex/aa}
\bibliography{./references}

\end{document}